# Implementation and evaluation of a dynamic contrast enhanced MR perfusion protocol for glioblastoma using a 0.35T MRI-Linac system

Danilo Maziero*[1,2], Ph.D., Gregory Azzam[1], M.D., Ph.D., Macarena de La Fuente[3], M.D., Radka Stoyanova[1], Ph.D., John Chetley Ford[1], Ph.D., Eric Albert Mellon[1], M.D., Ph.D.

[1]Department of Radiation Oncology, Sylvester Comprehensive Cancer Center, University of Miami Miller School of Medicine, Miami, FL, 33136, United States.
[2]Department of Radiation Medicine & Applied Sciences, UC San Diego Health, La Jolla, CA 92093, United States.
[3]Department of Neurology, Sylvester Comprehensive Cancer Center, University of Miami Miller School of Medicine, Miami, FL, 33136, United States.

Corresponding author:

Danilo Maziero, Ph.D.
Department of Radiation Medicine & Applied Sciences
UC San Diego Health
3855 Health Sciences Drive
La Jolla, CA 92093-0843, USA
(858) 246-1518
E-mail: dmaziero@health.ucsd.edu

**Abstract**

Purpose: MRI-linear accelerator (MRI-Linac) systems allow for daily tracking of MRI changes during radiotherapy (RT). Since one common MRI-Linac operates at 0.35T, there are efforts towards developing protocols at that field strength. In this study we demonstrate the implementation of a post-contrast 3DT1-weighted (3DT1w) and dynamic contrast enhancement (DCE) protocol to assess glioblastoma response to RT using a 0.35T MRI-Linac.

Methods and materials: The protocol implemented was used to acquire 3DT1w and DCE data from a flow phantom and two patients with glioblastoma (a responder and a non-responder) who underwent RT on a 0.35T-MRI-Linac. The detection of post-contrast enhanced volumes was evaluated by comparing the 3DT1w images from the 0.35T-MRI-Linac to images obtained using a 3T-standalone scanner. The DCE data were tested temporally and spatially using data from the flow phantom and patients. $K^{trans}$ maps were derived from DCE at three time points (a week before treatment–Pre RT, four weeks through treatment–Mid RT, and three weeks after treatment–Post RT) and were validated with patients' treatment outcomes.

Results: The 3D-T1 contrast enhancement volumes were visually and volumetrically similar (±0.6-3.6%) between 0.35T MRI-Linac and 3T. DCE images showed temporal stability, and associated $K^{trans}$ maps were consistent with patient response to treatment. On average, $K^{trans}$ values showed a 54% decrease and 8.6% increase for a responder and non-responder respectively when Pre RT and Mid RT images were compared.

Conclusion: Our findings support the feasibility of obtaining post-contrast 3DT1w and DCE data from patients with glioblastoma using a 0.35T MRI-Linac system.

# 1. Introduction

The Response Assessment in Neuro-Oncology (RANO) is often applied to measure glioblastoma response on MRI. These criteria primarily compare volumetric changes on post-contrast three-dimensional T1 weighted (3D-T1) and T2-weighted fluid-attenuated inversion recovery (T2-FLAIR) images[1]. MRIs are usually obtained one month before radiotherapy (RT) start and one month after treatment completion. This relatively long interval, approximately 3 months, does not permit early adaptation of RT for poorly responding tumors and reflects the uncertainty often seen in early post-RT MRI[2, 3]. In attempts to better define early response towards RT adaptation, studies have investigated MRI-derived physiological biomarkers such as tumor perfusion[4, 5], cell density[6], metabolism, and infiltration[7]. Such biomarkers have been reported to change during the course of RT and predict tumor response to treatment and patient outcome[8-10]. This physiologic information can be extracted by multiparametric-MRI (mpMRI) techniques[11, 12] such as dynamic enhancement contrast (DCE perfusion)[13], diffusion weighted imaging (DWI)[14] and magnetic resonance spectroscopy (MRS)[15].

The assessment of tumor response by mpMRI during the course of RT requires additional MRIs [16]. This can be challenging for many centers due to need for a standalone MRI, availability, scan coordination, and patient tolerance. The popularization of combination MRI-linear accelerator systems (MRI-Linac) allows for frequent mpMRI measurements without significantly disturbing the clinical routine. A previous DWI study detected treatment-related changes on the apparent diffusion coefficient (ADC) of head and neck tumors using a 0.35T MRI-guided tri-cobalt 60 system[17]. Another study implemented a quantitative MRI protocol[18], originally developed for 3T scanners, to obtain different relaxometry maps (i.e. R1, R2* and proton density) in patients with glioblastoma using a 0.35T MRI-Linac system[19]. Encouraging results for evaluating glioblastoma response during RT were also reported by a study that analyzed DCE data acquired every two weeks through the course of treatment[20]. However, the study was performed on a standalone 3T scanner, and its feasibility on the 0.35T MRI-Linac system has not yet been reported. In addition, there is a need to evaluate the ability of the 0.35T MRI-Linac to obtain post-contrast 3D-T1 images for adaptation of brain tumors based on anatomical changes[21].

In this study, we have implemented a T1 post-contrast and DCE protocol and evaluated the feasibility of providing perfusion-derived information using a 0.35T MRI-Linac. First, we evaluated the protocol's capability of measuring temporal changes due to the presence of contrast bolus using a perfusion phantom. Second, we evaluated the temporal stability of the DCE images obtained. Third, we evaluated the accuracy of contouring contrast enhanced volumes by comparing contours drawn on images from the MRI-Linac to contours drawn on images from a 3T scanner. Finally, we assessed the feasibility of estimating and applying parametric response mapping (PRM) based on $K^{trans}$ [22] to evaluate the response to RT of two patients with glioblastoma. This was done by utilizing data acquired one week before the treatment began (simulation), four weeks through treatment and four weeks after treatment completion on both 3T and 0.35T systems. Additionally, we share our experience regarding implementing the protocol on a radiotherapy service and highlight steps taken towards guaranteeing patient safety during the procedure.

## 2. Methods

### 2.1 MRI patient set-up

Images were acquired using a 0.35T MRI-Linac (ViewRay MRIdian, Mountain View, CA, USA). The acquisitions were obtained with the patients immobilized in a custom thermoplastic mask. At the time of writing there was no vendor-supplied dedicated head coil. Patients were imaged with the vendor-supplied head and neck anterior flexible coil array elements and torso posterior flexible array elements wrapped around the thermoplastic mask and baseplate (total 11 channels).

### 2.2 MRI acquisition

#### 2.2.1 Pre and post-contrast three-Dimensional T1 (3D-T1) weighted MRI.

A 3D gradient-recalled echo was used to acquire 36 axial slices of 2.5mm thickness, covering the entire brain, Repetition Time (TR)/Time to Echo (TE)= 40/4.84ms and a flip angle = 35º with a FOV=300x300x90mm$^3$. The reconstruction matrix was 210x210x36 pixels, resulting on a spatial resolution of 1.43x1.43x2.5mm$^3$. The total acquisition time for each 3D-T1 was 7 minutes and 30 seconds. These images were acquired using the 0.35T MRI-Linac system. The 3D-T1w images obtained using this protocol were compared to 3D-T1w images acquired using a MPRAGE sequence with 1x1x1mm$^3$ spatial resolution and TR/TE = 2000/2.54ms from a standalone Skyra 3T scanner (Siemens, Erlangen, Germany). Thirty-three slices of T2-FLAIR images with TR/TE = 9000/100ms and spatial resolution of 0.38x0.38x5mm$^3$ were also obtained using the 3T scanner. On the acquisitions using the 3T scanner, the patients were positioned supine headfirst, and a 32-channels head coil was used. The coil pads provided by SIEMENS were used to improve patient's immobilization during acquisitions using the 3T scanner.

#### 2.2.2 DCE MRI

A 3D gradient recalled echo sequence was used to acquire continuous volumes with 18 axial slices of the brain with the following parameters: TR/TE = 8.1/1.72ms, matrix = 100x100, FOV = 300x300x90mm$^3$, FA=12°, 18 slices and GRAPPA acceleration factor = 2 (phase-encoding direction, 24 reference lines). The acquisition time for each volume was 9.75s, totaling 11min22s (70 volumes) for the flow phantom experiments (section 2.2) and 7min19s (45 volumes) for the *in vivo* acquisitions.

2.3 Flow phantom experiment

An MRI compatible flow phantom (Shelley Medical Imaging Technologies, Toronto, Canada) was used for the *in vitro* portion of this study. The MRI-compatible parts of the phantom were set inside the MRI-Linac vault. The acrylic housing and the imaging plane (Fig. 1A) were placed on the MRI bed with the imaging plane being positioned at the isocenter. The non-MRI compatible pieces (i.e., peristaltic pump, valves and flow detectors) were set outside the Faraday cage but still inside of the MRI-Linac vault. This area is beyond the 5 Gauss line and the MRI-radio frequency sealing door, which was closed during our studies. The compatible and non-compatible MRI parts were connected by three 0.625 cm diameter and 20 m long hoses (here called Tubes 1, 2 and 3). Such connections were possible by passing the tubes through a connector path built in the front wall of the Faraday cage. Because our studies were done on MRI-Linac research mode, its accelerator was never active and therefore we did not need to seal the vault completely.

The pump was set to provide a continuous 0.5 Liter/minute flux of distilled water through the entire MRI acquisition of experiments 1 and 2 (sections 2.2.1 and 2.2.2). Two 250ml water bottles were positioned on the MRI bed, one beside Tube 1 and one beside Tube 3. This was done for improving the magnetization within the sensitive volume of the coils and to provide regions with constant signal through the experiments.

2.3.1    Experiment 1: Sensitivity to gadoteridol bolus

The objective of this experiment was to evaluate the sensitivity of our protocol to a concentration of gadoteridol similar to the one delivered to the tumor environment in DCE study. The typical concentration of gadoteridol administered intravenously to patients is 0.2 mL/kg (0.1 mmol/kg)[23]. The contrast is typically administered at a rate ranging from 2 to 4 mL/s [24, 25], and the time for the full administration is therefore dependent on patient's weight. Considering a typical resting heart rate of 60 bpm, and that a typical stroke volume contains approximately 60-75 mL of blood, it is reasonable to assume that with such administration rate, the blood pumped by the heart will have a ratio of 2 mL of gadoteridol per 70mL stroke volume of blood (~3% gadoteridol). Therefore, for this experiment we have injected a 1.5 mL bolus of gadoteridol into the 50 mL water (3% gadoteridol) in our system circulating at the same flux rate (0.5 L/minute).

The first bolus of gadolinium was injected into our system after continuously imaging for 2 minutes. After that we repeated injections of similar concentration every 30 seconds until the end of the acquisition. This was done to simulate the contrast accumulation commonly visualized in DCE data of patients with glioblastoma.

For this experiment, the output valves V2 and V3 (Fig. 1A) were set so the flux through V3 was 30% higher than the flux through V2.

### 2.3.2 Experiment 2: Gadoteridol bolus wash out

To evaluate the detection of the signal decrease related to the bolus of contrast leaving the area of interest, we have set the pump with the same flux parameters as the experiment described above (distilled water flux of 0.5 L/minute). However, we have assumed the higher value for the rate in which the contrast agent was administered (4 mL/s), therefore we have injected a bolus of 50 mL of distilled water with 6% of gadoteridol in volume into our system. The first bolus of gadolinium was injected into our system when the image acquisition began. The second bolus was injected 180s after the first.

For this experiment, the output valve V3 was kept closed until the second bolus of contrast was injected into the system. At that moment, the valve V3 was set so its flux was 30% higher than the flux through V2.

## 2.4 Patients' acquisitions

The patients were enrolled to a prospective non-interventional brain tumor imaging study that was approved by the University of Miami Institutional Review Board. The patients have provided verbal and written consent to participate in our study. As proof of principle, we present images from two patients with glioblastoma (for complete case descriptions please see results section 3.2), a 63-year-old male and a 57-year-old female. Patient 1 underwent contrast MRI scans on the 0.35T MRI-Linac system at three time points, during simulation a week before undergoing chemoradiation treatment (Pre-RT), four weeks after treatment start (Mid-RT) and 4 weeks after (Post-RT) radiotherapy completion. The images acquired with the MRI-Linac system were obtained in between 24-48 hours after the acquisitions using the standalone MRI scanner for the Mid and Post RT time points. The images from the Pre RT encounter were obtained on the MRI-Linac approximately eight hours after the acquisitions using the standalone MRI scanner. Patient 2 completed Pre and Mid-RT scans only as she died one week before the Post-RT follow up. The images acquired by the MRI-Linac system were obtained in between 24-48 hours after the acquisitions using the standalone MRI scanner.

Patients were treated for 30 fractions to a cumulative dose of 60 Gy over 6 weeks with all fractions given on the 0.35T MRI-Linac by step-and-shoot intensity modulated RT. The initial phase delivered 46 Gy in 23 fractions to the enhancing tumor and resection bed plus a 2 cm clinical target volume (CTV) and a 3 mm (PTV) planning target volume. At the beginning of week 5, the Mid

RT scans were used to plan a boost of 14 Gy in 7 fractions to the week 5 tumor and resection bed plus a 5 mm CTV and 3 mm PTV.

2.5 Patient positioning and safety for contrast administration

Once in the patient's treatment/imaging position (section 2.1), the patient's previously placed intravenous (IV) access was connected to the MRI-compatible injector (Bracco diagnostics, New Jersey, USA). IV access was wrapped around the patient's thumb and taped to secure its position to avoid the IV from being dislodged in case of abrupt movements. Finally, the patients were positioned with the center of their heads at isocenter.

The gantry of the Linac was set to 300 degrees, which was found empirically to decrease susceptibility artefacts while acquiring MRI data. For the perfusion acquisitions, the MRI-Linac was used in research mode, and therefore the linear accelerator was blocked and could not be active during the acquisitions. The vault door was opened to facilitate patient access in the unlikely case of a gadolinium reaction, while the Faraday cage door was closed. To increase the magnetic field homogeneity at the isocenter, the manual shimming tool was applied twice following the acquisition of a 3D localizer and field of view optimization for maximum coverage of the tumor and brain.

The protocol used for obtaining the images from the patients consisted of a pre-gadoteridol administration 3D-T1 (as described by section 2.1.2), 45 volumes of a 3D-GRE (section 2.1.3) and a post gadoteridol 3D-T1. The total acquisition time for the three sequences was approximately 23 minutes.

A 0.1 mmol/kg bolus of gadoteridol was administered with a 2 mL/s rate followed by a 30 mL saline solution administered at the same rate. Four baseline points of DCE were acquired before the contrast administration.

### 2.6 In vitro data processing and analysis

The data from the in vitro experiments were imported, analyzed, and visualized using MATLAB 2020a (MathWorks, Natick, MA, USA). In order to facilitate visualization, we have created three regions of interest, one for each tube, considering the center axial slice of the volume and the time courses from the voxels within each ROI were averaged generating three time courses.

### 2.7 In vivo MRI data pre-processing

All images acquired from each patient were co-registered to the post-contrast 3D-T1 obtained on the week before the treatment start (Pre RT) using MIM (MIM Software Inc., Cleveland, OH). Following the contouring of the regions of interest (see section 2.6), the images were imported to MATLAB 2020a (The MathWorks, Natick, USA) and motion correction of the dynamic data was applied referenced to the first volume time point using SPM 12 (http://www.fil.ion.ucl.ac.uk/spm/software/spm12).

### 2.8 Spatial analysis of anatomical images

#### 2.8.1 Tumor and regions of interest delineation

Lesions designated as "Tumor" were manually contoured on the post-contrast enhancement 3D-T1 weighted images using MIM. An extended "Peritumor" region of interest (ROI) was created by expanding the margins of the tumor contours by 0.8 cm. A third region of interest ($Exp_{2cm}$) was created by expanding the margins of the Tumor contours by 2 cm. Only tissues associated with tumor, gray and white matter were considered for the Peritumor ROIs, therefore regions containing cerebrospinal fluid, skull and scalp were excluded from the extended contours. This step was done for 3D-T1 images acquired using both MRI-Linac system and standalone MRI.

The different ROIs (Tumor, Peritumor and $Exp_{2cm}$) from images obtained from each scanner were compared by estimating their spatial overlap, which was further normalized by the volume of the ROI contoured on the 3T images. Such percentage overlap was obtained using the following equation:

$$Overlap_{ROI(C)} [\%] = 100 \times \frac{Volume\ (ROI(C)_{3T} \cap ROI(C)_{0.35T})}{Volume\ (ROI(C)_{3T})} \qquad \text{Eq. 1}$$

Where, $ROI(C)_{3T}$ and $ROI(C)_{0.35T}$ are the contours for a specific region of interest C (Tumor, Peritumor or $Exp_{2cm}$) obtained from images acquired using the 3T standalone and the 0.35T MRI-Linac systems, respectively. The operator $\cap$ was used to calculate the spatial intersection among $ROI(C)_{3T}$ and $ROI(C)_{0.35T}$.

Two additional ROIs were created by contouring the 3D-T1 weighted images acquired on the MRI-Linac system. These ROIs contained voxels from vessels or white matter regions contralateral and posterior to the tumor. These ROIs were labeled Artery and Control, respectively.

### 2.8.2 Atlas based volumetric analysis

Voxel-based morphometry (VBM) was performed with CAT12[26] and SPM toolboxes. The pipeline included gray matter (GM), white matter (WM), and cerebrospinal Fluid (CSF) tissues segmentation, followed by non-linear spatial registration to the MNI (Montreal Neurologic Institute) template, bias correction of intensity non-uniformities, and modulation (to account to the amount of volume changes after spatial normalization). The GM volumes were estimated in the subject native space using the Neuromorphometric atlas (Neuromoorphometrics, Inc), by individually applying the inverse Jacobian matrix estimated during normalization to the atlas reference image and, successively, to the atlas itself. This analysis was focused on structures that were not neighboring the cortical and subcortical lesions. Please see supplementary material Tables S1 and S2 for the list of the structures considered for each patient.

2.9 Temporal stability and ROI time course analysis

The DCE data obtained from each timepoint were evaluated by calculating their temporal signal-to-noise ratio (tSNR). The tSNR is the ratio of the mean signal and standard deviation of each voxel within the brain considering its time evolution through the entire acquisition. This metric provides a visual and quantitative evaluation of the data's temporal stability [27], and it is useful for highlighting areas showing large signal changes (i.e., artefact-related such as dropouts). In this study, we have calculated the tSNR after excluding the first eight time points of our dynamic acquisition. This was done to avoid including the sharp signal rise caused by the gadoteridol administration. Additionally, we have applied a mask to exclude the voxels representing the ventricles and CSF from this evaluation. The differences among tSNR from different acquisitions were evaluated by ANOVA and corrected for multiple comparisons ($p_{Bonferroni}<0.05$) when necessary.

The voxels within the ROIs Peritumor, Artery and Control were averaged creating three distinct time courses. This was done to verify the temporal behavior of different regions of the brain due to the gadoteridol administration.

2.10 Quantitative pharmacokinetic parameter ($K^{trans}$) mapping

The DCE data were spatially smoothed using a gaussian filter with FWHM=5mm. The datasets were composed by 45 time points and the first four were used as pre-contrast temporal changes baseline. The $K^{trans}$ maps were calculated based on the single compartment model proposed by Tofts and Kermode[22]. Pre-gadoteridol injection T1 values were not available for the patients analyzed in this study. Therefore, we have considered pre T1=700 ms for the volume of the tumor based on other datasets obtained by our group[16] and a previous relaxometry study performed at 0.35T[19]. The R1 relativity after injection was assumed to be 4.7 $s^{-1}$ based on values reported by Tofts and Kermode[22] and adjusted for the 0.35T magnetic field strength according to previous work [28, 29]. The arterial input function used was population-based as previously described[30].

The differences between the $K^{trans}$ values from different acquisitions were evaluated by ANOVA and corrected for multiple comparisons when necessary. The threshold for statistically significant differences was chosen as $p_{Bonferroni}<0.05$.

### 2.11 Parametric response evaluation

The $K^{trans}$ maps were limited from 0 to 0.3 $min^{-1}$ and submitted to a voxel-wise parametric response evaluation[9]. The $K^{trans}$ parametric response map was obtained by calculating the difference between $K^{trans}$ (e.g. $\Delta K^{trans}$ = Mid RT $K^{trans}$ – Pre RT $K^{trans}$) for each voxel within the tumor on Pre and Mid treatment data. Therefore, a voxel initially found within the contrast enhanced volume on Pre RT images would be excluded if it was found outside the contrast enhanced volume on Mid RT images. A preestablished threshold based on the 95% confidence interval of all values of each map was considered to determine that a voxel showed $\Delta K^{trans}$ decrease (responding to treatment) or increase (failing to treatment).

# 3. Results

## 3.1 In vitro acquisitions

### 3.1.1 Experiments 1 and 2

The signal changes due to a gadoteridol bolus were visible for each one of the ROIs within each tube (Fig. 1B). While the comparison of volumes obtained before (Fig. 1B, Vol 1) and after (Fig. 1B, Vol 70) the bolus injection showed negligible averaged signal changes for the control water ROIs W1 (1.15% +/- 8%) and W2 (0.45% +/- 9.7%), they showed an average increase of, 192% +/- 10%, 252% +/- 19% and 238% +/- 24% for tubes 1, 2 and 3 respectively (Fig. 1C, left hand side). Our experimental setup allowed us to detect a temporal delay between the bolus input (Tube 1) and output of the system (Tubes 2 and 3). This was observed more clearly on the data from experiment 2 (Fig. 1D). Finally, the results from experiment 2 also showed that doubling the gadoteridol volume concentration (from 3% to 6%) used during experiment 1 also doubled the signal increase within the tubes' ROIs (Fig. 1C versus Fig. 1D).

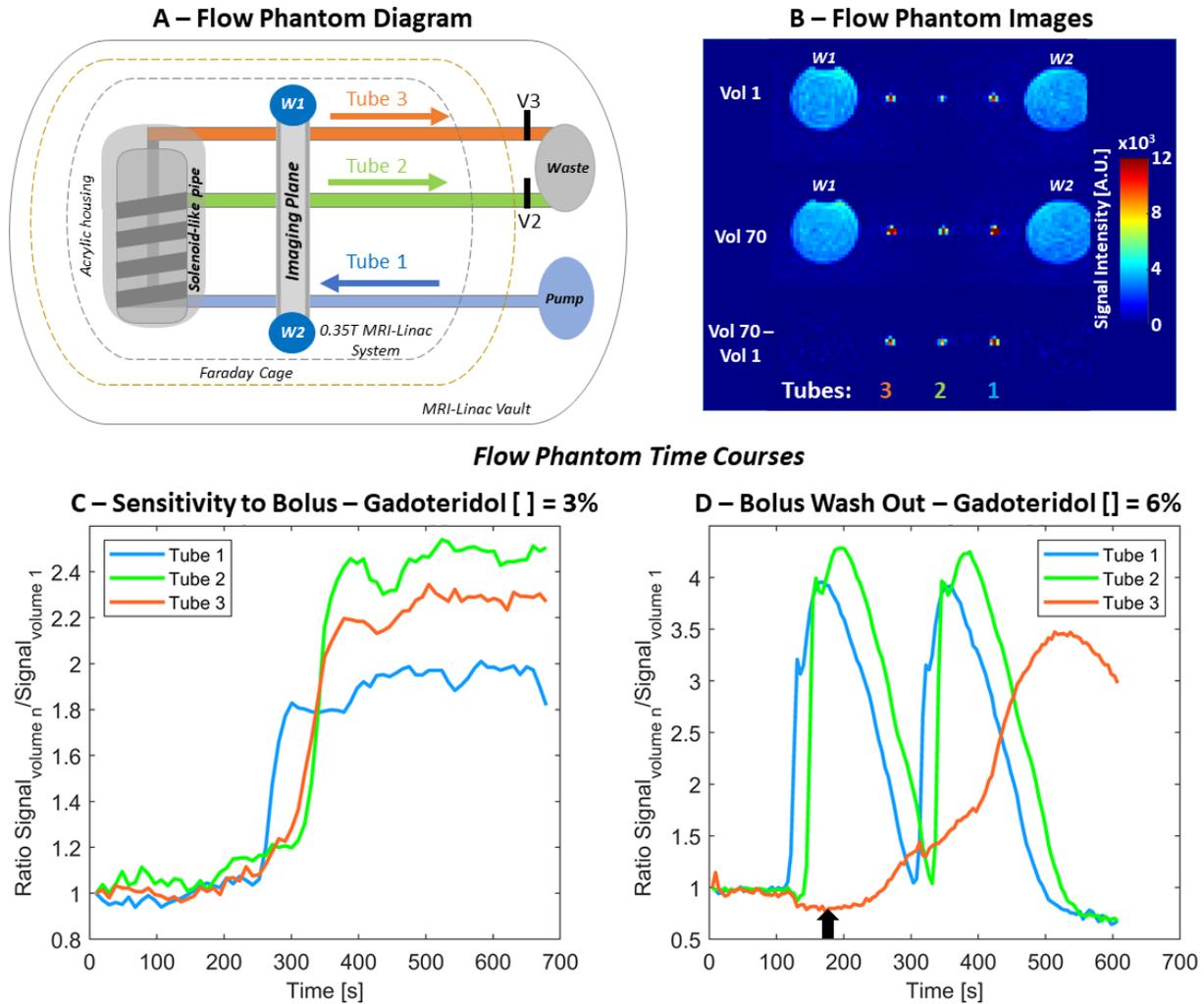

Figure 1: Evaluating the temporal capability of detecting a bolus of gadoteridol using a flow phantom. The flow phantom is composed of a peristaltic pump, three tubes, an acrylic housing for a solenoid-like pipe and an imaging plane (A). Valves V2 and V3 are used to control the output flux through Tubes 2 and 3, respectively. (B) Axial slices of flow phantom images. The same axial slice is exhibited for volumes 1 (Vol 1, at 9.75 seconds, top row), 70 (Vol 70, at 682.5 seconds, middle row) and the pixel-by-pixel subtraction Vol 70 – Vol 1 (bottom row) for the data acquired during the sensitivity to bolus experiment. The control water volumes (W1 and W2) are not influenced by the injection of gadoteridol in the system, and therefore disappear due to the Vol 70 – Vol 1 subtraction. (C) Signal change due to a 3% in volume injection of gadoteridol for tubes 1 (blue), 2 (green) and 3 (orange). (D) Signal change due to a 6% in volume injection of gadoteridol for tubes 1 (blue), 2 (green) and 3 (orange). Signal change differences between tubes 1, 2 and 3 are mainly due to partial volume and susceptibility effects. The delay seen in tube's 3 (Fig. 1D) signal time course, in comparison to tubes 1 and 2, is because V3 (Fig. 1A) was kept closed until the second bolus of contrast was injected into the system (Fig. 1D, black arrow). Therefore, there was no flux through Tube 3 before that.

3.2 Patient acquisitions

Patient 1 is a 64-year-old male who presented with complex partial seizures leading to brain imaging demonstrating a 6.8 cm enhancing mass in the right frontal lobe. He underwent stereotactic core needle biopsy demonstrating IDH-1 wildtype glioblastoma. Insufficient viable tissue was available for O(6)-methylguanine-DNA methyltransferase (MGMT) testing. Pre-RT MRI was performed 6 days prior to start of RT and temozolomide, mid-RT MRI was performed at fraction 19, and post-RT MRI was performed 3 weeks after RT. Patient had possible progression of disease on MRI 5 months status post (s/p) RT while on adjuvant temozolomide chemotherapy and bevacizumab was initiated with stable disease at last follow-up 6 months s/p RT.

The MRI data obtained mid-RT were interpreted by a radiologist that is not involved in this study and was reported as following: "IMPRESSION: Interval decreased size of the large right frontal lobe/basal ganglia heterogeneously enhancing necrotic mass, with overall improving associated surrounding confluent T2/FLAIR hyperintense signal as detailed above. Improved mass effect on the right lateral and third ventricles. Interval improvement/resolution of midline shift. Persistent elevated CBV in the solid peripheral components. Continued follow-up recommended."

The Post-RT MRI data interpretation was: "IMPRESSION: Overall decreased size of the heterogeneously peripherally enhancing multilobulated necrotic mass centered in the right frontal lobe/basal ganglia with involvement of the body of the corpus callosum. Persistent internal areas of restricted diffusion with elevated CBV along the periphery noted."

Patient 2 is a 57 year old female who presented with cognitive decline, generalized weakness, and incontinence. MRI demonstrated 7.0 cm enhancing mass centered in the anterior corpus callosum. Stereotactic core needle biopsy demonstrated IDH-1 wildtype glioblastoma. Tumor was extensively necrotic and MGMT testing was inconclusive due to extensive necrosis. Pre-RT MRI was performed 8 days prior to start of RT and temozolomide and mid-RT MRI was performed at fraction 19. Unfortunately, the patient declined cognitively two weeks after RT and died three weeks after RT prior to post-RT MRI.

The MRI data obtained mid-RT were interpreted by a radiologist that is not involved in this study and was reported as following: "IMPRESSION: Large necrotic mass centered within the corpus callosum, centrally necrotic with irregular peripheral enhancement and elevated CBV corresponding to enhancing components of the lesion. No significant change compared to prior study."

We have not detected any imaging discrepancies that could be associated with crosstalk between the two gadolinium administrations.

### 3.2.1 Spatial analysis of anatomical images

#### 3.2.1.1 Tumor and region of interest delineation

The post gadoteridol administration 3D-T1 images (Fig. 2A) from patient 1 (responder) and patient 2 (non-responder) using a 0.35T MRI-Linac (top row) and 3T standalone MRI (bottom row) showed similar contrast enhancing volumes when compared to each other. This was observed for images acquired Pre (red), Mid (green) and Post RT (blue, Post RT images from patient 2 were not available). The contrast enhanced volume for patient 1 images from pre, mid and post therapy were, respectively, 3.6, 3.2 and 3.5% larger for contours drawn on images acquired with the 0.35T MRI-Linac in comparison to contours drawn on images from the 3T stand-alone scanner (Fig. 2B). The contrast enhanced volume for patient 2 images from pre and mid therapy were, respectively, 0.6 and 2.4% smaller for contours drawn on images acquired with the 0.35T MRI-Linac in comparison to contours drawn on images from the 3T stand-alone scanner. As described in section 3.2, Patient 1 showed a response to treatment, where the contrast enhanced volume detected using the 0.35T MRI-Linac decreased in 27.6 and 38.4% from Pre to Mid and from Pre to Post treatment images, respectively. Although a 14.8% decrease in the overall volume of the enhancement was also detected from pre to mid RT images from patient 2, the contrast enhancement became thicker in regions distributed across the entire volume. Finally, the effect of not applying a coil sensitivity correction while using parallel imaging strategies is observed on the pre-RT 3D-T1 (Fig. 2A, top, left corner). The pre and post contrast 3D-T1 from patient's 1 Pre RT appointment were the only images that such correction was not applied.

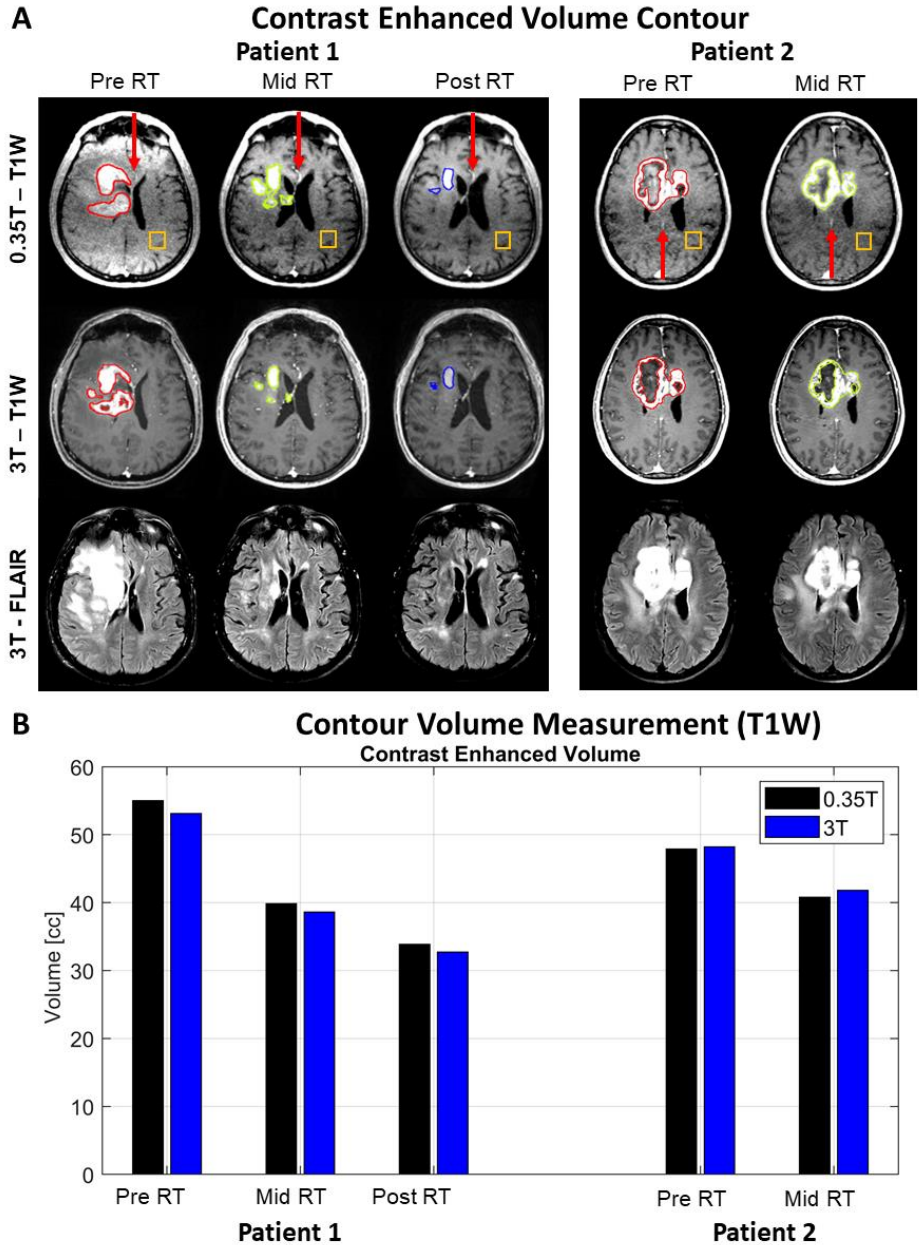

Figure 2. Post gadoteridol T1 weighted images from 0.35T MRI-Linac and standalone 3T scanner from two patients with glioblastoma. The contours (fig. 2A) highlight the contrast enhanced volume on images acquired before radiotherapy treatment (red), fourth week of treatment (green) and one month post treatment (blue) using a 0.35T MRI-Linac (top row) and a standalone 3T MRI scanner (middle row). The T2-FLAIR images (bottom row) obtained from both patients using a 3T scanner are displayed as complementary information about the tumor. Patient 2 did not respond to the treatment and died before the post-treatment follow up. The volume of each contour for images acquired on the 0.35T MRI-Linac (black) and clinical 3T scanner (blue) are also presented for the available time points (fig. 2B). The red arrows and the orange boxes on the images obtained with the 0.35T MRI-Linac indicate the ROIs used for the averaged artery and control time courses, respectively.

The ROIs contoured on images from the 3T and 0.35T showed spatial overlaps ranging from 81.2 to 91.7% for the ROI Tumor, from 95.2 to 97.6% for the ROI Peritumor and from 96.1 to 98.3% for the ROI Exp2cm, respectively (Table 1).

| Table 1. Overlap $_{ROI(C)}$ [%] | | | |
|---|---|---|---|
| **Treatment Time Point** | **Tumor** | **Peritumor** | **Exp$_{2cm}$** |
| *Patient 1 - Responder* | | | |
| **Pre RT** | 87.6 | 95.4 | 97.1 |
| **Mid RT** | 84.5 | 95.3 | 96.1 |
| **Post RT** | 88.9 | 95.2 | 96.6 |
| *Patient 2 – Non Responder* | | | |
| **Pre RT** | 91.7 | 97.6 | 98.3 |
| **Mid RT** | 81.2 | 95.9 | 97.4 |
| **Post RT** | N/A | N/A | N/A |

3.2.1.2 Atlas based volumetric analysis

The atlas based volumetric analysis of different brain structures using T1 weighted images from the 0.35T MRI-Linac and the 3T standalone systems presented coefficients of determination ($R^2$) ranging from 0.96 to 0.99 for patient 1 (Fig. 3A) and from 0.98 to 0.99 for patient 2 (Fig. 3B) when compared.

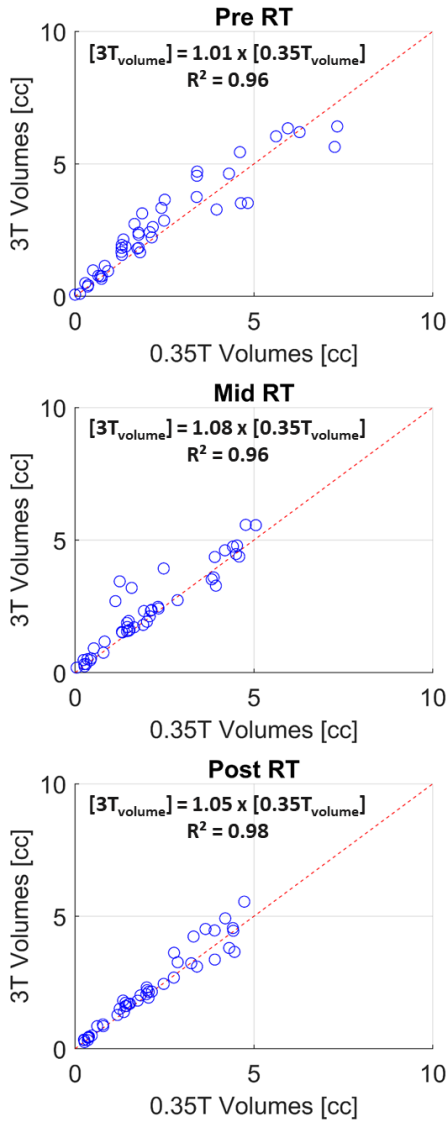
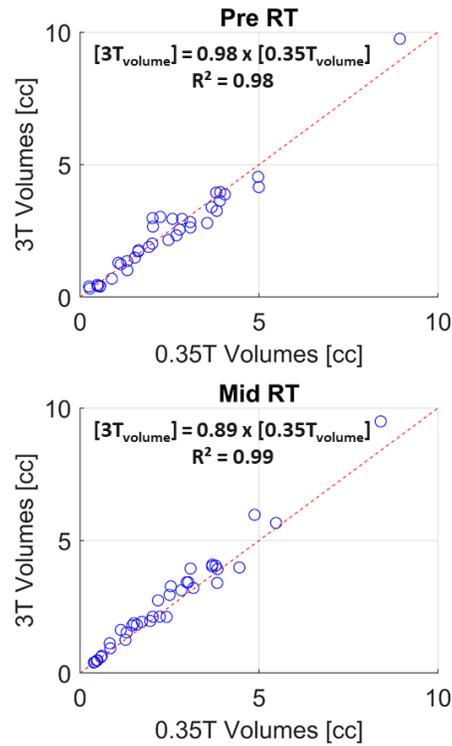

Figure 3. Atlas based volumetric analysis of different brain structures based on T1 weighted images from patients 1 (A) and 2 (B). For each scatter plot the x-axis and y-axis contain the volumes in cc of each structure analyzed using images from the 0.35T MRI-Linac and the 3T standalone systems, respectively. The equation resulting from the linear fit considering the correlation among all pairs of points and the $R^2$ coefficient for such fit are shown in each one of the panels. From top to bottom the panels show the results using images from Pre, Mid and Post RT for patient 1 and from Pre and Mid RT for patient 2.

### 3.2.2 Temporal stability analysis

The temporal stability of the DCE data was evaluated using the tSNR metric. The tSNR maps indicate a homogeneous distribution for brain voxels (Fig. 4A, top row) across the entire volume for patients 1 and 2. The tSNR for Pre RT, Mid RT and Post RT DCE data (Fig. 4A, second row) from patient 1 were 10.4 +/- 2.3, 12.2 +/- 3.7 and 11.6 +/- 3.0, respectively. For patient 2, the tSNR were 13.4 +/- 4.5 and 14.5 +/- 4.9 for data from Pre RT and Mid RT, respectively.

The temporal evolution of the averaged signal from the ROIs Peritumor, Artery and Control exhibited three different behaviors as expected (Fig. 4B). For both patients we observed that the Peritumor ROI time courses (Fig. 4B, top row) showed sharp increases due to the gadoteridol administration followed by a plateau (patient 1) or slow further increase (patient 2) of the signal, for data obtained Pre (Fig. 4B, top panel, red), Mid (Fig. 4B, top panel, green) and Post RT (Fig. 4B, top panel, blue, patient 1 only). The signal changes for each volume are relative to the first time point, therefore presented in percentage. The signal change after the initial slope was 11.8, 12.1 and 13.9% for patient's 1 Pre, Mid and Post RT data, respectively, and 7.95 and 7.2% for patient's 2 Pre and Mid RT data, respectively. The artery ROI time courses (Fig. 4B, mid row) showed a signal sharp increase due to the gadoteridol administration followed by a slow decrease for both patients' data obtained from Pre (Fig. 4B, mid panel, red), Mid (Fig. 4B, mid panel, green) and Post RT (Fig. 4B, mid, blue). The peak of signal change following the initial slope was 19.2, 34.9 and 23.6% (Fig. 4B, second row) for patient's 1 Pre, Mid and Post RT data, respectively, and 58.9 and 48.0% for patient's 2 Pre and Mid RT data, respectively. The average signal change within the Control ROI (Fig. 2A, orange box) was 3.85 +/- 1.32%, 0.87 +/- 0.94% and 2.57 +/- 1.11% for patient's 1 data from Pre, Mid and Post RT visits, respectively. For patient's 2 Pre and

Mid RT data, the average signal change within the Control ROI was 2 +/-1.42% and -0.64 +/- 1.21% respectively

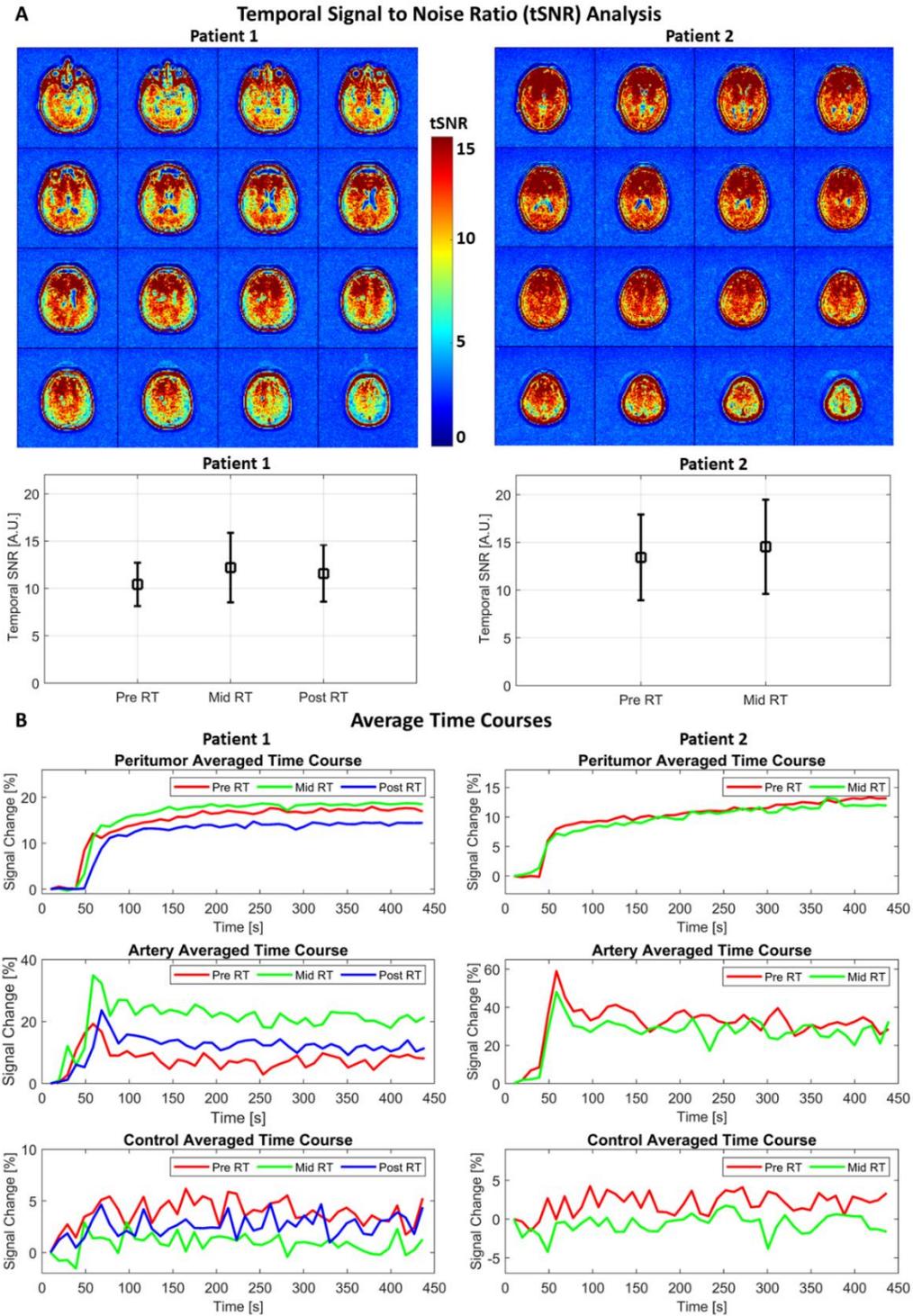

Figure 4. Temporal analysis of the images obtained for tumor perfusion evaluation using the dynamic contrast enhancement technique. The temporal stability of the data was evaluated for each set of dynamic volumes acquired (Fig. 4A) using the temporal SNR metric. The tSNR of sixteen slices from Mid RT volumes (top row) are presented for patients 1 (left-hand side) and 2 (right-hand side). The average tSNR and its standard deviation are presented (Fig. 4A, bottom row) for Pre RT, Mid RT and Post RT for patient 1 (left-hand side) and Pre RT and Mid RT for patient 2 (right-hand side). The averaged temporal courses (Fig. 4B) for the peritumor region (top panel), an artery (middle panel) and a control region (bottom panel) are presented for patients 1 (left hand-side) and 2 (right hand side). The time courses from Pre RT (red), Mid RT (green) and Post RT (blue) are presented for patient 1. The Post RT data were not available for patient 2.

### 3.2.3 $K^{trans}$ maps and functional parametric response analysis

The perfusion data from patient 1 (responder) showed a progressive decrease in $K^{trans}$ values from Pre to Mid and Post RT (Fig. 5A, from top to bottom panel, respectively). The decreases were observed in most of the peritumor region (red contour), with one exception that is highlighted by the yellow arrow. The averaged $K^{trans}$ from patient's 1 (responder) maps decreased in 54% from pre to Mid RT (from 0.1471 to 0.08 min$^{-1}$, $p_{Bonferroni}<0.05$) and decreased additional 16% from Mid to post RT maps (0.081 to 0.068 min$^{-1}$, $p_{Bonferroni}<0.05$), respectively. Despite of the small decrease in the volume on the contrast enhanced region of patient 2 (Fig. 2B), the perfusion data showed an 8.6% increase of averaged $K^{trans}$ values from Pre to Mid RT maps (0.0924 to 0.1004 min$^{-1}$, $p<10^{-4}$) for this non-responder patient (Fig. 5B).

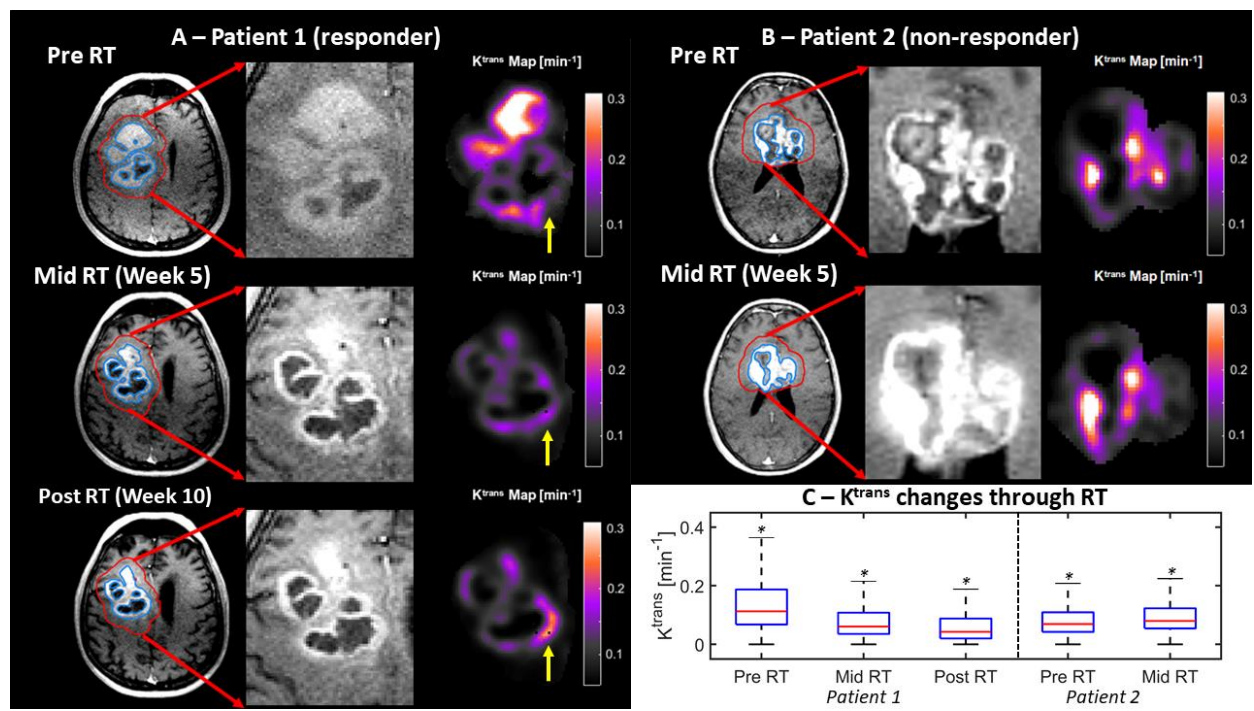

Figure 5. Perfusion changes over the course of chemoradiation treatment. In each panel in A and B an axial slice of the post contrast 3D-T1 (with two contours: tumor – blue and peritumor – red), a zoom into the peritumor and the $K^{trans}$ map for that region are presented from the left to the right-hand side, respectively. A - From top to bottom, provides information from data acquired before treatment start (Pre RT), during treatment (Mid RT, from week 5) and post treatment (Post RT, week 10) from patient 1 (responder). The yellow arrow highlights a region that potentially could have failed treatment. Figure B, from top to bottom shows the data acquired before treatment start (Pre RT) and during treatment (Mid RT, from week 5) from patient 2 (non-responder). C – The summary of the $K^{trans}$ maps changes through treatment for a responder (patient 1) and a non-responder (patient 2). The * highlights statistically significant differences found by ANOVA (a Bonferroni correction was applied to data from patient 1).

The total number of voxels within the ROI Peritumor (red circle in Fig. 5A and B top row) were 5452 (245.3 cc) and 4850 (218.25 cc) for Pre RT DCE data from patients 1 and 2, respectively. A total of 3168 (142.6 cc) and 896 (40.32 cc) voxels showed either increase or decrease in $K^{trans}$ values when data from Pre and Mid RT were compared for patients 1 and 2, respectively. The histogram for the $K^{trans}$ values of voxels within the peritumor ROI presented a considerable shift downwards (Fig. 6A, left-hand side) when patient's 1 (responder) data from Pre and Mid RT were compared. On the other hand, a shift slightly upwards was verified when the $K^{trans}$ histograms from patient's 2 (non-responder) Pre and Mid RT data were compared (Fig. 6A, right-hand side). The parametric response mapping evaluation of patient's 1 (responder) data (Fig. 6B and 6C) showed that 80 and 20% of the voxels presented decrease and increase on $K^{trans}$ values, respectively. The same analysis of patient's 2 (non-responder) data showed that 36.0 and 64.0% of the voxels presented $K^{trans}$ values decrease and increase, respectively.

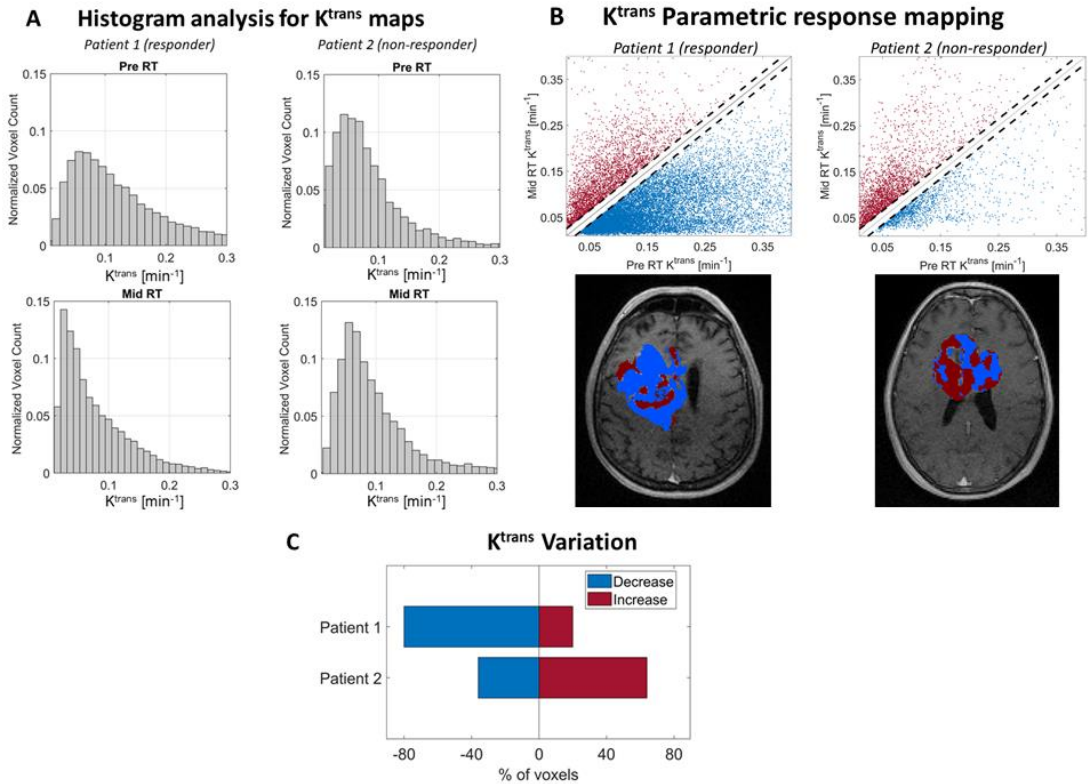

Figure 6. $K^{trans}$ histogram and parametric response mapping analysis. A – The normalized histograms for $K^{trans}$ values from patients 1 (left hand side) and 2 (right hand side) are shown for Pre RT (top row) and Mid RT (bottom row) data. An overall summary of each voxel response to treatment is presented by fig. B (top row) for patients 1 (left hand side) and 2 (right hand side). The $K^{trans}$ values obtained from Pre RT data (x-axis) and Mid RT data (y-axis) for each voxel are presented as a scatter plot. The number of points above or below the diagonal black dashed lines represent voxels with increased (red) or decreased (blue) $K^{trans}$ respectively, meaning a negative and positive response to treatment. One representative slice is shown to illustrate the voxels that are responding (blue) or not (red) to treatment of a responder (patient 1, left hand side) and non-responder (patient 2, right hand side) patients. Figure C summarizes the percentage of voxels that responded (blue) or not (red) to treatment in patients 1 (responder) and 2 (non-responder). The characterization of a voxel as responder or non-responder is based on the difference between $K^{trans}$ values from Pre RT and Mid RT data. The voxels showing at least 0.015 $min^{-1}$ decrease from Pre

to Mid RT data are considered responders (blue), on the other hand voxels showing at least 0.015 min$^{-1}$ increase from Pre to Mid RT data are considered non-responders (red).

## 4. Discussion

We have implemented and evaluated a protocol to acquire MRI-perfusion using a 0.35T MRI-Linac system. The DCE data obtained from the *in vitro* experiments 1 and 2 showed that our protocol has spatial and temporal sensitivity (Fig. 1, B to C) to detect signal changes caused by a gadoteridol bolus of similar concentration to that used in patients. The data resulting from both experiments allowed for the visualization of signal changes within a voxel with a volume = 0.045 cc (3x3x5 mm$^3$). This volume represents less than 0.1% of the contrast enhanced volume of patients 1 and 2 on Pre RT images (Fig. 2B). The capability for accurately detecting contrast enhanced volumes on a 0.35T MRI-Linac was also supported by the visual and volumetric similarities observed on contours drawn over post contrast 3D-T1 images from 0.35T MRI-Linac and 3T standalone scanner (Fig. 2). Temporal SNR and ROI time courses showed stability in phantoms and patients with expected curves for tumor, artery, and control brain (Fig. 4B). The application of a compartmental model for analyzing the tumor's pharmacokinetic profile showed results for K$^{trans}$ maps in concordance to previously reported for patients with glioblastoma[20]. While an overall K$^{trans}$ decrease through RT correlated with response to treatment for patient 1 (Fig. 5A), an increase of K$^{trans}$ indicated treatment failure for patient 2 (Fig. 5B). The parametric response analysis further allowed for the detection of portions of the tumor potentially responding and failing RT (Fig. 6) which might be useful in the future for planning RT boost treatments.

Imaging at low magnetic field strength raises concerns about decreased SNR in comparison to higher fields (i.e., 1.5 and 3T) and the T1 relaxivity of gadolinium chelates. The desire for higher SNR is driven by the possibility of increasing spatial resolution or decreasing acquisition times.

However, the increased SNR and spatial resolution are not always translated linearly to diagnostic improvement when images from low and high field scanners are compared. The visual and volumetric analysis performed for the post-contrast 3D-T1 images obtained from 0.35 and 3T magnetic field strengths showed evidence in agreement with that (Fig. 2). Additionally, we verified a strong correlation ($R^2>0.96$) among the volumes of different brain structures when comparing results obtained from images acquired with the 0.35T MRI-Linac and 3T standalone systems that were submitted to an atlas based volumetric analysis (Fig. 3). It is important to highlight that the in-plane resolution of our 3D-T1 images acquired on the 0.35T MRI-Linac were 1.43x1.43mm$^2$ (vs 1x1mm$^2$ of images from the 3T scanner), therefore having the largest difference on the slice thickness 2.5mm for the images from the 0.35T (vs 1mm for images from the 3T). Although we verified a comparable diagnostic power (0.35T vs 3T) for the two patients evaluated in this study, it is important to note that the tumors studied here are relatively large and that significant differences may be detected for tumors of smaller volumes (i.e. brain metastasis). The evaluation of other tumor types is therefore a topic for further studies.

The capability of imaging the accumulation of gadolinium chelates, such as gadoteridol, in human brain and spine tumors have been previously reported for a range of magnet strengths (0.2-1.5T)[31]. As such, many of our dynamic contrast enhanced-derived parameters were derived from data acquired using a 0.5T scanner[22]. More thorough discussion about advantages and opportunities using low field MRI is available[32]. In this study, we optimized the imaging parameters for the DCE acquisition to match the guidelines provided by the quantitative imaging biomarkers alliance from the Radiological Society of North America (RSNA) as closely as possible[33]. We were able to match their suggested parameters for temporal resolution (<10s) and slice thickness (5mm). The TR of our protocol was slightly longer than the suggested (8.1 vs 7ms) with a 1x1mm$^2$ loss on the in-plane slice resolution (3x3 vs 2x2 mm$^2$) consequently to the decreased signal-to-noise ratio of imaging on a 0.35T magnet. Despite such limitation, we were able to observe the evolution of tumor enhanced volumes through the course of RT for a responder and a non-responder patient (Fig. 5).

Dynamic MRI datasets are often affected by a variety of artefacts that may decrease their temporal stability and cause an increase in false positives and negatives of quantitative measurements [34, 35]. In this study, we evaluated the temporal stability of the DCE datasets by analyzing their temporal signal-to-noise ratio (tSNR) and signal changes due to a gadoteridol bolus. Such analysis showed that the temporal stability was homogenous across the brain for all sessions and patients (Fig. 4A), supporting the data capability to be further analyzed by quantitative metrics. Besides that, we observed that signals from ROIs associated with tumor, artery, and background noise (Fig. 4) exhibited temporal behaviors similar to previously reported[36, 37]. The availability of data through the course of RT allowed us to observe changes on the time courses of ROIs containing multiple sources of signals. For example, the time course of the ROI 'artery' from patient 1 Pre RT DCE data (Fig. 4B) showed a smaller peak due to the bolus of gadoteridol than the time courses of the same ROI from Mid and Post RT DCE. We hypothesize that such difference is explained by an overlap of the tumor tissue with the vessel being observed (Fig. 2A), which decreased with the shrinkage of the tumor due to its response to radiotherapy.

The availability of MRI-Linac systems to the clinic provides a potential for observing the response of different tumors to RT more frequently through the course of treatment. Previous studies showed the potential of detecting tumor diffusivity[17] and relaxometry[19] changes using a 0.35T MRI-Linac during RT and that such changes are potentially correlated to the tumor response to treatment. Another study showed that changes in tumor volume can also be detected through the course of treatment by investigating T2 weighted images in a series of 14 patients with glioblastoma being treated on a 0.35T MRI-Linac[38]. Here, we showed that the tumor dynamics due to contrast administration can also be studied using a similar system and that patient response to treatment may be observed by analyzing $K^{trans}$ maps (Fig. 5). Additionally, we provide evidence that PRM analysis of $K^{trans}$ maps is feasible at 0.35T and may provide relevant clinical information about responding and non-responding patients (Fig. 6). For example, although patient 1 (responder) showed a continuous decrease of mean $K^{trans}$ values through the course of treatment and a downwards shift on individual voxels $K^{trans}$ values, it presented an increase of such values for one region of the tumor (Fig. 5A, yellow arrow). The long-term outcome for this region, and its potential correlation to treatment failure needs to be carefully addressed by a longitudinal study with a larger cohort of similar cases and it will be topic for a further study. Nevertheless, this example illustrates the potential use of DCE in a 0.35T MRI-Linac for detecting portions of the tumor that are resistant to treatment early on the course of RT (week 5 in this case) and could provide with an option for treatment adaptation. It is important to note that the ideal frequency for DCE data acquisition and therefore gadolinium administration is a controversial topic in literature and is still being discussed [11]. Previous studies have successfully obtained DCE datasets every other week without reporting any reaction to the administration of gadolinium to patients with glioblastoma[20]. We believe that this administration frequency might be close to the achievable at the moment. The justification for administering gadolinium more frequently than that might be difficult as there is little data to support its safety/benefit.

The implementation and optimization of sequences and protocols commonly available for relatively high magnetic field diagnostic MRI scanners (e.g., 1.5 and 3 T) can be challenging at 0.35T due to a decrease of SNR. Such decrease may be compensated to a certain extent by trading off on spatial and temporal resolution. In this study, such tradeoff was necessary, resulting 3D-T1w images with spatial resolution of 1.43x1.43x2.5 mm$^3$ at 0.35 T, compared to 1mm$^3$ isotropic from a 3 T scanner. Despite such limitation, images obtained by our protocol quantified the contrast enhanced volume similarly to images from a 3T scanner. Additionally, it provided post contrast 3D-T1w images visually comparable (Fig. 2) to those obtained by the diagnostic scanner, supporting the findings of previous studies[32, 39, 40]. Besides that, our results suggest that using post-contrast T1 images from a 0.35T MRI-Linac system may aid on adaptive RT as tumor changes are detected through the course of therapy. This was illustrated by the high spatial overlap (>95.2%, Table 1) between volumes commonly used as PTV (e.g. 0.8 to 2cm expansion over the contrast enhanced volume detected) that resulted from images acquired with the 0.35T and 3T systems. A second limitation of our study was the temporal resolution of the DCE data (9.75 s). Although such parameter was faster than the recommended by the RSNA (<10 s)[41], a previous study showed that the accuracy of estimating the arterial input function (AIF) may be affected for similar temporal resolution[36]. The impact of this limitation was seen in our results (Fig. 4B, middle panel), where the peaks observed for the AIF were not as sharp as showed by other studies that acquired DCE data with fast temporal resolution (e.g., <5 s)[36, 42]. It is important to note that in this study we used a population-based AIF[30] instead of patient-specific. Therefore, the limitation related to the relatively long temporal resolution and potential errors associated to detecting such function was not carried over to the calculation of the $K^{trans}$ maps. The sensitivity of $K^{trans}$ maps has been reported to be affected by multi variables related to the AIF selection and preprocessing and is still a topic of discussion in literature[30, 43-46]. The evaluation of these variables and their impact on DCE modeling and $K^{trans}$ sensitivity would require a larger group of patients than the one presented here, and therefore will be evaluated in a further study.

Another limitation of our study was the time availability for acquiring the 3D-T1w and DCE data. The pre and post gadoteridol 3D-T1w and DCE data were obtained in approximately 25 minutes. Because of that, we were not able to obtain pre-contrast relaxometry data and had to assume a fixed T1 for $K^{trans}$ calculation. A similar strategy was assumed by previous studies[22, 47]. We believe that the implementation of modern strategies for acquiring dynamic MRI data (e.g. non-cartesian k-space trajectories and parallel imaging)[40, 48, 49] on a 0.35T scanner will aid improving the temporal and spatial capabilities of post-contrast 3D-T1w and DCE data acquisition and will be investigated in a further study. If resolution can be sufficiently improved, MRI simulation post-contrast can become available at 0.35 T for multiple applications including glioblastoma and brain metastases.

## 5. Conclusion

MRI-Linac systems provide an opportunity to evaluate tumor response frequently during RT. In this work, we have implemented a protocol for detecting gadoteridol-based contrast enhancement and dynamic contrast enhancement of glioblastoma using a 0.35T MRI-Linac system. We have evaluated our protocol by acquiring in vitro and in vivo data and verified that it is able to detect signal changes due to a bolus of gadoteridol in a phantom and in two patients with glioblastoma (a responder and a non-responder). In our results we have verified that tumor enhancement volume is virtually the same by comparing measurements on post 3D-T1 images obtained from 0.35 and 3T scanners. Additionally, we have shown that tumor perfusion can be evaluated in a 0.35T MRI-Linac and that changes on $K^{trans}$ maps during treatment can be detected and may be associated to tumor changes due to RT and patient outcome. Such changes were shown to be detectable as early as the fifth week of treatment, which highlights the potential of this technique to be used for adaptive radiotherapy of patients with glioblastoma using a 0.35T MRI-Linac system.

## 6. Acknowledgements

This work was supported by the National Center for Advancing Translational Sciences of the National Institutes of Health (NIH) under Award Number UL1TR002736, and the National Cancer Institute of the NIH under Award Number R37CA262510 and K12CA226330 as well as Sylvester Comprehensive Cancer Center intramural support, and the Dwoskin Cancer Research Fund. The content is solely the responsibility of the authors and does not necessarily represent the official views of the National Institutes of Health or other supporters. The authors would like to thank Dr. Yu-Cherng Chang for sharing the MATLAB codes used as base for analyzing the perfusion data and Dr. Brunno M. Campos for helping with the atlas-based volumetric analysis.